\documentclass[conference]{IEEEtran}
\IEEEoverridecommandlockouts
\usepackage{url}
\usepackage{hyperref}
\usepackage{amsmath,amssymb,amsfonts}
\usepackage{algorithmic}
\usepackage{graphicx}
\usepackage{textcomp}
\usepackage{xcolor}
\usepackage{multirow}
\usepackage{float}
\def\BibTeX{{\rm B\kern-.05em{\sc i\kern-.025em b}\kern-.08em
    T\kern-.1667em\lower.7ex\hbox{E}\kern-.125emX}}
\begin{document}
\title{Designing the Drive: Enhancing User Experience through Adaptive Interfaces in Autonomous Vehicles}
\author{
\IEEEauthorblockN{Reeteesha Roy}
\textit{VIT Bhopal University} \\
reeteesharoy@gmail.com
}

\maketitle

\begin{abstract}
With the recent development and integration of autonomous vehicles (AVs) in transportation systems of the modern world, the emphasis on customizing user interfaces to optimize the overall user experience has been growing expediently. Therefore, understanding user needs and preferences is essential to the acceptance and trust of these technologies as they continue to grow in prevalence. This paper addresses the implementation of HCI principles in the personalization of interfaces to improve safety, security, and usability for the users.This paper explores the way that personalized interfaces can be devised to increase user engagement and satisfaction through various HCI strategies such as adaptive design, multi-modal interaction, and user feedback mechanisms. Moreover, this paper puts emphasis on factors of transparency and user control in the design of an interface; hence, allowing users to design or modify their experience could foster an increase in trust in autonomous systems. In so doing, this research touches on the quite influential role HCI will play in this future scenario of autonomous vehicles while designing to ensure relevance to the diverse needs of users while maintaining high standards of safety and security. Discussing various HCI strategies such as adaptive design, multi-modal interaction, and feedback mechanisms to the user, this paper demonstrates how personalized interfaces can enhance significantly both user engagement and satisfaction. Transparency and user control also in designing an interface are further discussed, pointing out the need for a prerequisite condition of enabling the user to take control of their experience as a state of trust in autonomous systems. In summary, this paper points out the role of HCI in the development of autonomous vehicles and addresses numerous needs with respect to those enforced safety and security standards.
\end{abstract}

\vspace{5pt}


\vspace{10pt}

\section{\textbf{Introduction}}
Autonomous vehicles, or more commonly known as self-driving or driver less cars, are the pinnacle of autonomous technology. Their working principle is based on the combination of multiple advanced and complex technologies like sensors, actuators, cameras, GPS, and voice-activated AI that allows these vehicles to sense their environment, make acceptable decisions, and with less to no human interference\cite{1}. They're designed to follow the traffic rules and therefore reduce the probability of road accidents caused by human error. They prioritize safety and accessibility for the user and help in reducing overall environmental degradation by lower fuel consumption and reduced carbon emissions.
Personalization, with reference to technology, refers to the tailoring of digital systems to fit the taste of each user\cite{2}. This involves analyzing user data, such as past interactions, search histories, and demographic information, to create customized content, recommendations, and interface layouts. By employing algorithms that adapt to user preferences, technology can enhance engagement and satisfaction, making interactions more relevant and meaningful. For instance, streaming services use personalization to suggest shows and movies based on viewing habits, while e-commerce platforms recommend products tailored to user interests.Moreover, effective personalization also involves providing users with control over their experiences. Users should have the ability to adjust settings, customize preferences, and opt-in or opt-out of data collection for personalization purposes\cite{3}. This transparency not only fosters trust but also empowers users to shape their interactions with technology in a way that aligns with their comfort levels and values. By balancing tailored experiences with user agency, technology can create a more enjoyable and efficient user journey, ultimately enhancing overall satisfaction and loyalty.

HCI has evolved since its invention in 1970 to fill the gap between humans and machines, with a focus on usability and accessibility of the system as well as interaction design. Historically, HCI has been studied with the aim of making digital systems more efficient and making them easy to use for humans via GUIs, keyboards, and eventually touchscreens. As technology evolved, researchers at HCI have been more focused on implementing interaction design beyond personal computers, leading to the innovation of virtual reality, mobile computing, and automotive systems.

In the autonomous vehicular industry, HCI principles were applied to ensure the safety of dashboard controls and navigation aids, reducing the cognitive load of the drivers and ensuring that they could interact with the system freely.Autonomous vehicles have since developed into a trendy and relevant field through means of self-learning artificial intelligence models, machine learning, and sensor technologies, including newly integrated systems of driver-assist technologies like cruise control, lane keeping, using voice recognition, and real-time feedback\cite{4}. As they gained popularity and momentum, HCI started focusing on making interaction between complex, intelligent, and independent systems and humans more and more manageable. HCI researcher Merat et al. (2014) focused on the seamless transition of vehicle control between humans and the autonomous system. Now-a -days, HCI aims to adapt systems to build trust and enhance comfort for the users.

In Section(II), the problem statement and the objectives of this review are discussed, in Section(III) the methodology of the review is described, in Section(IV) the key concepts from the objects are
reviewed and in Section(V) the findings from the review are presented.

\section{\textbf{Problem formulation}}

\subsection {Human Computer Interaction}
Human-Computer Interaction is a multi-disciplinary discipline focusing on the design and use of computer technology, especially the interaction between humans (users) and machines. HCI, therefore, in the context of autonomous vehicles would comprise intuitive interfaces that could provide optimum user experience. HCI ensures that users can communicate with the intricate vehicle's systems such as navigation or driving control or even infotainment without ever feeling they are being cognitively burdened. The goal is making the task less complex while still retaining a high degree of usability and accessibility. Another related aspect of HCI for AVs is multimodal interaction including touchscreens and voice commands with gestures, in which real-time feedback and clear communication of vehicle actions and boundaries are provided \cite{5}. This makes interaction much smoother because the user feels much more in control with confidence in the capabilities of the vehicle. HCI increases the trusts, comfort, and security within AV systems because it promotes user-centered design.

\subsection {Autonomous Vehicles}
Autonomous vehicles, often known as self-driving cars or driverless vehicles, are one of the leading edge autonomous technologies related to transport\cite{6}. All of these advanced technologies such as sensors, cameras, radar systems, and GPS in combination with artificial intelligence allow it to drive and run without the interference of humans. AVs are programmed to perceive the surroundings, make decisions in real time, and follow set traffic regulations to bring under control the risks of accidents that normally prevail due to human nature. These vehicles therefore bring about the minimization of traffic congestion and harmful emissions through efficient fuel consumption and ecologically friendly driving patterns. Their end is always achieving safe, efficient, and environmentally friendly transportation systems. Being able to be combined with IoT and intelligent infrastructures, AVs lead various industries to pave their way into a smarter urbanization into smart cities.

\subsection {Personalization}
The system, interface, or service is tailored to the individual's own choice, behavior, and need. For example, personalization for an autonomous vehicle is more rooted in personal experience development through customized driving styles, seat adjustment, climate setting, and infotainment based on the profile of the user.The data thus collected from user interactions, past behaviors, and preferences are used in creating a unique interface that responds dynamically to each and every user.Personalization may thus introduce comfort along with the challenge of user engagement in the actual meaningful experience\cite{7}. For example, an AV would vary its driving speed based on a user's history of liking either conservative or aggressive driving styles. Cabin temperature, lighting, or entertainment may be varied on their time of day, or even to lift the user's mood. That establishes trust by giving users choices, such as opting in or out of features and data collection, which will give transparency and allow the user to make their choices.

\subsection {User Experience(UX) }

 User Experience in the case of AVs is needed because it is no longer sufficient to have an efficient working system; users should feel comfortable and enjoy the experience\cite{8}. Starting with the design of seats and layout of dashboards to digital interactions related to navigation, voice control, or real-time feedback, every part indeed influences UX. Personalization will form the core of tailoring the system to the user's preferences, such as routes, entertainment options, and even mood settings\cite{9}. A well-designed UX will promote satisfaction but also improve safety if, indeed, users are kept focused and informed when necessary. Ultimately, a seamless UX will enhance trust and acceptance in AV technology because of clear, understandable feedback.
As the count of autonomously driven vehicles is constantly on the rise, personal interfaces based on trust, safety, and satisfaction are warranted\cite{10}. Real-life complex systems through which the working of AVs is understood to have decidedly hinged on AI and IoT do not meet the requirements of its users if such personal interfaces are not incorporated into the usage process. Users vary in what they need from the kind of drive or comfort settings and interaction with their AVs, so if these needs are not met, this causes a disconnection between the user and the vehicle.

Besides the appeal to the driver's comfort, safety and transparency of AV systems are also needed. A user should be informed of what is going on but not in control at least for risky situations such as lane change or passing through traffic flow\cite{11}.Therefore, the question then is exactly when to pass control so as not to communicate the intent of sacrificing safety on the altar of multimodal convenience yet retain user's trust. Matters have to be handled by developing the interface so that it brings out HCI principles for adaptive, user-centered designs being responsive to the trends of individual preference while ethical, privacy, and security considerations are put into place. A specialized, transparent AV system that meets the expectation of safety is aimed at here.

\newpage

\textbf{\textbf{The objectives of this review are stated as follows:}}
\begin{itemize}
\item Personalized driver emulation and embedding of the same in the cognitive engine that is at the core of UI.
\item Training of the UI so that the driver develops trust. Algorithm development for the same.
\item Lane change and other complicated procedures should be a part of AV capability.
\item It should be enabled with sensor IoT and voice recognition capabilities so that it can get intelligent transportation system infrastructures like roadside units at its fog computing and other assistive service through modern 5G/6G and beyond wireless connectivity.
\item Driver's understanding of safe driving should be emulated through an HCI based computational learning system, and the sensors should play the role of digital skin, which alerts and offers feedback about the dynamic environment.
\end{itemize}

\section{\textbf{Methodology}}

\begin{figure}[H]
    \centering
    \includegraphics[width=1\linewidth]{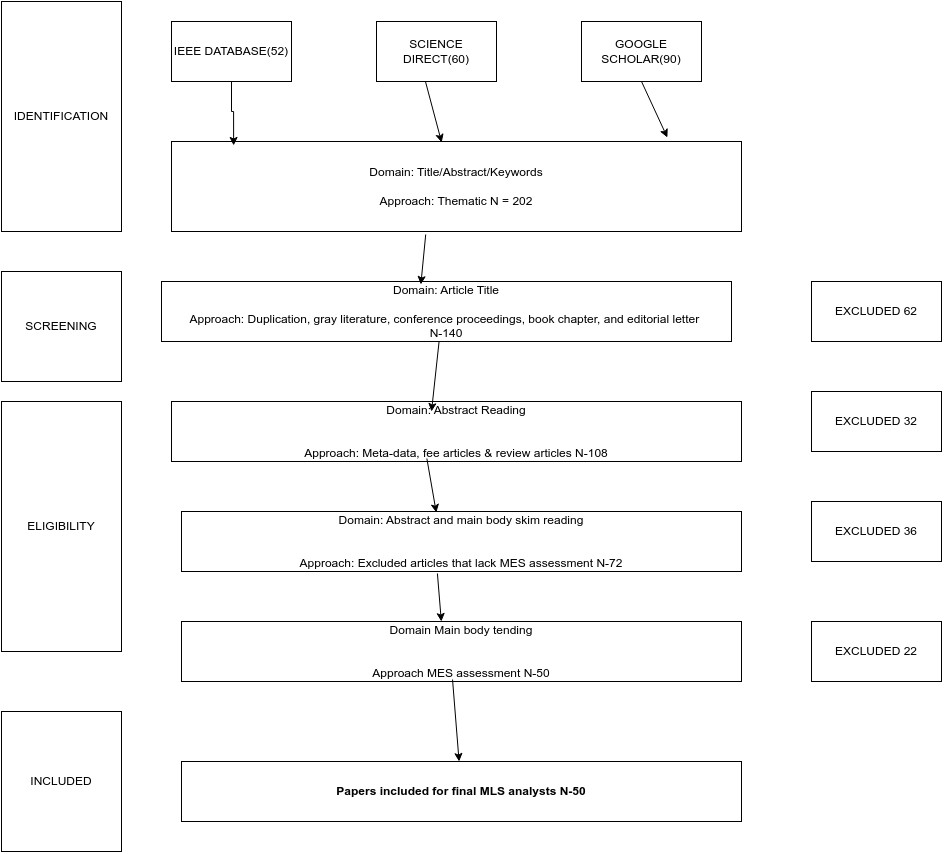}
    \caption{Systematic Review Process}
    \label{fig:Methodology}
\end{figure}
Observed in Fig. 1, the methodology explains the systematic
review process intending to identify 202 articles starting from
the source Google Scholar(90),  IEEE Database(52) and Science Direct(60). During the screening phase 62 articles were removed due to duplicate titles, gray literature, and editorial letters remaining with 140 articles eligible to be abstracted. In the reading abstracts stage, another 32 articles were removed finding fee based barriers, irrelevant metadata, and review articles decreasing the pool to 108 articles. The skim reading of abstracts and main body excluded 36 articles for lacking Model/Method Evaluation Strategy (MES) assessments narrowing down numbers to 72 articles.

After going through the entire main body, 22 more articles were removed by the helps of MES assessment against the number of selected 50 papers were included in comprehensive MES assessment.                 

\section{\textbf{Systematic Review of Ideas}}
\subsection{Principles of HCD in Autonomous Vehicles}

\subsubsection{What is HCD?}

Human-Centered Design (HCD) is a problem-solving approach to product development as well as interface design that focuses on the user preferences and user experience. HCD emphasizes the creation of viable solutions that are usable, accessible, and meaningful to the user rather than just technological advancement alone\cite{12}.
The elements of HCD in the autonomous industry are as follows:

Empathy for user: Research, interviews, and taking feedback to understand the need and pain points of the users.
Iterative Design: Developing a prototype to improve based on the expectation of users.
Involve Users: Involve users at every step in the process, incorporate their feedback into better designs.
Problem Solving : Deal with the actual pain points, which the users are facing, through practical solutions rather than more technical aspects.
Holistic Approach: In idea formulation, the user experience should be considered at different levels, from emotional to the cognitive levels.
Accessibility and Usability: The product to be accessible to various groups, that is, different abilities at different skill levels.

Users must be comfortable with and in control of the use of the system. Trust forms a crucial objective to this design since, in trust, the users are leaving their lives on the roads to the system. To build such trust, HCD ensures the interface forms a key point of contact between the vehicle and the system, which must be responsive to different human needs\cite{13}.

\subsubsection{Fundamentals of HCD}

The fundamentals of the user-centric approach in designing personalized interfaces for autonomous vehicles (AVs) are to get to understand and satisfy these special needs and preferences along with those behaviors unique to one's users, thus making the system adaptable to individual differences\cite{14}.
The first step toward personalization of autonomous vehicles is understanding user interaction with such vehicles.Improvement of the system can be done by collecting data through interviews, surveys, and user observations in the analysis of pain points, preferences, and behaviors and ascertaining overall user requirements\cite{15}. For example, some might prefer conservative driving, while others want faster or more dynamic driving.

AV systems also have personalized interfaces, which can vary with each user's specific context\cite{16}. For example, preferences in terms of the user preferences may depend on factors such as time of day, destination, or even the mood of the user at the present moment. Context-aware personalization is hence used to automatically adapt settings related to seat positions, cabin climate control, or entertainment options based on such precise conditions.

Autonomous vehicles are used to customize driving styles as per the user. A senior person or a new one would like to have a careful driving style, whereas a practicing person may need quicker response time during traffic.A user-centric approach is straightforward in terms of being as simple as possible and easy to understand in terms of feedback methods\cite{17}. They appear in visual displays and voice commands that relate to individual experiences.This leads to establishing trust with the technology as well as engaging users with the technology, making users feel more comfortable and satisfied with the product. AVs, having an understanding of the user's preferences, offer safety features catering to one's individual needs.In conclusion, a user-centric approach in AVs ensures the delivery of personalized, adaptive interfaces to enhance the overall user experience.

\subsubsection{Personalization of UX}

UX personalization involves crafting digital experiences in such a way that better caters to the needs, preferences, and behaviors of users at an individual level\cite{18}. In this regard, gathering data on user interactions provides actionable insights for adaptive changes in content, layout, and functionality to change. An interface would study trends in previous usage and make customized recommendations, thus enhancing engagement and satisfaction.

Effective personalization also needs mechanisms of feedback that may be provided by the users for their preferences and settings in order to maintain comfort\cite{19}. Giving the users control over personalization develops trust and persuades them to engage more with the system. This, eventually, leads to more intuitive and relevant digital interactions, hence improving the overall user journey.
\subsection{Factors}
\subsubsection{User Preference and Trust}
User preferences have a great impact on the acceptance and usability of autonomous vehicles\cite{20}. An effective design in HCI, therefore, needs to focus on understanding the needs of the individual user and tailoring experiences for them. Of course, trust will be paramount, where users need to believe in the decisions of a vehicle and safety measures taken. HCI can foster trust by providing transparent feedback about the action of the vehicle, real-time situational awareness, and clearly communicating the limitations and capabilities of the system.

\subsubsection{Multi Modal Approach}
Such a multi-modal autonomous vehicle enhances the interaction between the user and vehicle through the use of voice commands, touchscreens, and gestures\cite{21}. In this case, flexibility is allowed in how the user selects a preferred mode of interaction, considering context and personal comfort. According to HCI principles, transitions between modes should be seamless: users must be allowed to change from one input mode to another without losing control and context.

\subsubsection{Features}
The features in an autonomous vehicle are designed with HCI in order to provide a friendly environment for the users. Advanced navigation, personalized dashboards, and entertaining functions will make the journey quite interactive\cite{22}.Concepts of HCI also ensure that safety feature designs inform users about upcoming hazards or the decisions of the vehicle in advance through alerts and mechanisms of feedback. These features promote great satisfaction among users and instill confidence in the general experience of driving an autonomous vehicle due to usability and accessibility.

\subsection{Personalization in AV Interface: Components}
\subsubsection{GUI}
A graphical user interface, or GUI, is the computer, handheld device, and other electronics' visual representation of communication with the user that relies on heavy utilization of icons, buttons, and menus and allows users to intuitively interact with and navigate without relying on text-based input\cite{23}.
Graphical user interfaces on the autonomous vehicle enrich the passenger experience through real-time route navigation and display of traffic conditions, along with estimated arrival times. Besides that, the passengers can enter the destination through an easy-to-operate touchscreen; simply adjust the settings for climate control and seating. GUIs also provide critical statuses of the vehicle, which include speed and the battery level, with safety notifications of obstacles surrounding them\cite{24}. They also support infotainment features for passengers to listen to music and watch videos. User profiles on GUIs allow personalized settings, hence assuring a smooth and entertaining atmosphere for the passengers, building trust and comfort during the journey.

\subsubsection{Role of AI and ML}
Personalization of user interfaces for autonomous vehicle involves automated understanding of the environment around the vehicle\cite{25}. It includes recognition and identification of obstacles which appears in the form of other vehicles and roadway topology, incorporation of understanding of existence of other moving objects around itself while performing maneuverings like lane change or change of speed and direction. It should also be able to respond effectively and correctly to change of roadside traffic control signaling system in real-time.
In a smart city environment is essential for orchestration of society 5.0. The realization of the same is being done through various next generation Information, control, and communication technologies which as a whole offers us the layout of industry 4.0\cite{26}. As factories will be graduating up to smart factories producing customized products at a large scale, intelligent transportation infrastructure in domains like roadway transportation, railways and waterways, air and space travel will start putting its own footprint. As a consequence of the same, roadside units or RSU’s equipped with limited local cache memory, computing power, sensors, and wireless transceivers will appear which enabled by AI / ML supported next generation 5G/6G and beyond wireless connectivity will also play important role for these autonomous vehicles in securing local information, path planning based on traffic congestion etc\cite{27}.

Cognitive computing which includes various AI and ML related techniques will help develop personalized driver profiles that interacts with this multidimensional complex environment\cite{28}. It will help develop the necessary driver trust about the vehicle and offer a driving experience which they are comfortable and familiar with.

\subsubsection{Sensor Techonologies and IoT}
One of the important innovation of last few decades is the sensors and sensor network technology. Sensors of various sizes (Crossbow motes have small sensors in dimension, while salinity measuring seaside sensors can be somewhat bulky and big) appeared with various capabilities like gas sensing, underwater and under the surface monitoring, light intensity monitoring, vibration monitoring etc\cite{29}. based on various basic technologies like piezo electric effect, chemical reactions and others which enables the ICT systems to develop a digital skin to keep track of environmental changes. Naturally these technologies will have a great impact in providing environmental information in a dynamic manner to train the cognitive engine which will support the personalized user interface.

\begin{figure} [H]
    \centering
    \includegraphics[width=1\linewidth]{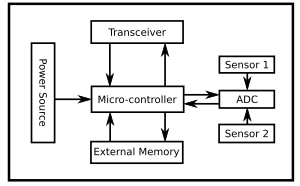}
    \caption{Typical Architecture of Sensor Node\cite{30}}
    \label{fig:Sensor Node Diagram}
\end{figure}

\cite{31}Next stage of development is sensor and wireless communication capability enabled individually addressable devices or smart appliances which can be connected to a local communication network which in turn can be connected to the greater internet. It helps support the concept of smart factories and smart home.These devices enabled by protocols like LoRa has supported a great deal of new innovation in smart factory realization\cite{32}. It is expected that this revolutionary technology will be integrated in realization of personalized user interfaces of autonomous vehicles which will allow the driver to get advanced information about the planned trajectory and can establish Inter vehicular communication to have a safer driving experience.  

\subsubsection{Voice Recognition}

Voice recognition in driver-less vehicles is an integral HCI component that enables natural language processing between passengers and vehicle systems to make easy communication possible\cite{33}. This means that the users can use voice input to command, ask questions, and control functions such as navigation, climate, and entertainment. Utilizing context-aware processing, the voice recognition system focuses on intent towards users and can cope with ambiguous requests, as well as tolerate various accents and speech patterns to make usability and accessibility.That is, it is developing a hands-free interaction method that reduces cognitive load and distraction, thereby improving the overall user experience in autonomous vehicles \cite{34}.

\subsection{Challenges}

\subsubsection{Data Privacy and Security}
Data protection and security in the realm of autonomous vehicles: these systems collect vast amounts of personal information on location, preference, and behavior. These imply that such unauthorized access and breaches are contained through stringent measures in data protection to ensure user confidentiality. Data privacy and security concerns in autonomous vehicles make HCI designs employ strong encryption methods in transmitting and storing data to protect against unauthorized access of user information. Privacy settings that allow users to tailor their choice about the data that is collected and with whom it will be shared need to be clearly embedded in user interfaces. Transparency features might include a variety of things, from alerts when data is in use to requests for consent. This could make users more confident and careful in decisions involving their own privacy.

\subsubsection{Ethical Considerations}
The ethics of autonomous vehicles: Algorithms deciding the first critical issue in autonomous vehicle design would indeed to be ethical, especially the algorithms in which decisions are taken\cite{36}. HCI comes out with the ethical view of how the vehicle likes to choose first safety and what they decide is likely to cause harm to people or to other elements by their actions. It deals with transparent systems that mirror the ethical frameworks and values of societies in such a manner that users can feel and trust the decisions taken by an independent system.

\subsubsection{Hyper-Personalization}
While hyper-personalization of services and interfaces improves user experience, it also raises serious concerns regarding data usage and the potential for autonomous users\cite{37}. In HCI, there are designers who weigh the advantages of personalization against possible intrusion into privacy or over dependence on technology. These could be some feedback mechanisms from the user himself through surveys or preference sliders providing input on desired levels of personalization. Second, personalization could be implemented using machine learning algorithms that respect user boundaries and preferences about privacy so that it enhances the experience without undermining user autonomy and choice about disclosure of personal information.

\subsubsection{Automation and User Control Balancing}
Balancing automation with user control can be achieved by creating intuitive interfaces and updating users in real-time on the action of the autonomous vehicle\cite{38}. HCI can enhance this through the implementation of embedded systems for visual and auditory feedback that will notify users of the state of a vehicle and possible hazards. Manual override options, easily accessible with a touch or voice command, further allow users to take control when desired, enhancing their sense of agency and confidence in the system.

\subsection{Case Studies In Recent Years}

\subsubsection{Case Study 1}
Lane Change Prediction in Urban Driving Simulations\cite{39}

Purpose: Predict lane changing by machine and human decisions in a driving simulator with HUD interface using Unity3D.

The experimental method consisted of two test scenarios:
1. Front vehicle impact The machine showed the lane change more slowly than the human, with host vehicle travelling at 60 km/h and front vehicle at 50 km/h.
2. Leading Vehicle Change Left: Computer would have easily anticipated the lane change at an earlier time, but human had hoped for a faster left vehicle to pass at 70 km/h.
Results:The humans' choices in both scenarios were safer and indicated that the machines are still not quite hitting the mark about decisions in complex conditions through intuition.
Conclusion: Lane-changing machine learning models should improve their decisions to better approach real-world conditions that human lane-changing decisions would assume.
\begin{figure} [H]
    \centering
    \includegraphics[width=1\linewidth]{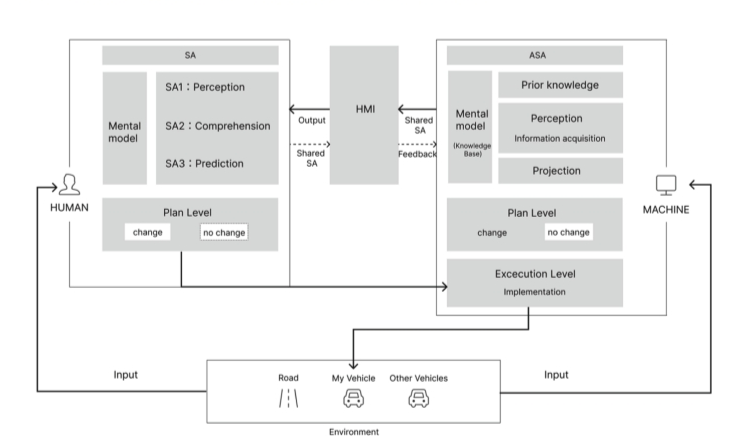}
    \caption{Team SA Relations within the Human-Machine System mentioned in the study}
    \label{fig:Team SA Relations within the Human-Machine System mentioned in the study}
\end{figure}
As we have observed they have designed personalized UI incorporated experimental test bed and got encouraging results on overtaking and lane change scenario.
\begin{figure} [H]
    \centering
    \includegraphics[width=1\linewidth]{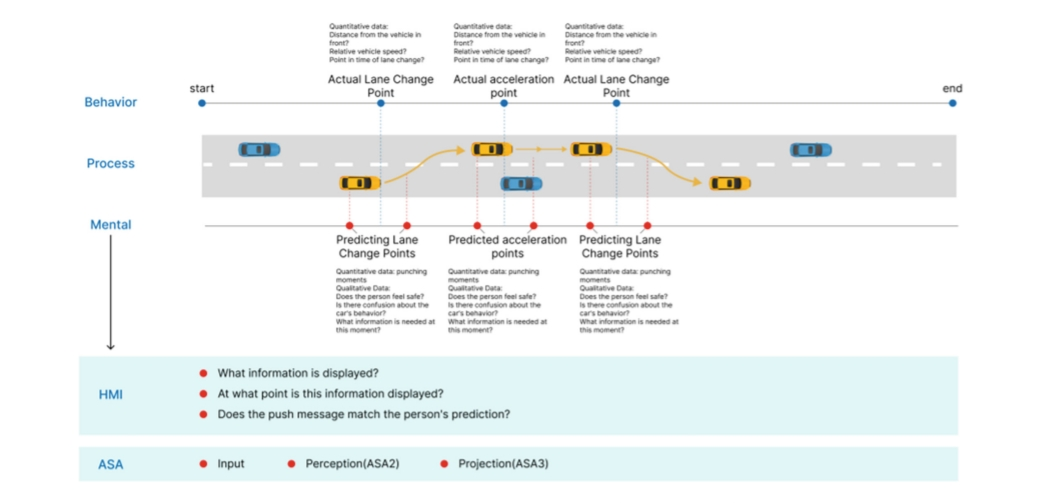}
    \caption{Lane Change Scenario}
    \label{fig:Lane Change Scenario}
\end{figure}
\subsubsection{Case Study 2}
Understanding User Trust in Autonomous Vehicles (AVs)\cite{40}

Objective:
This attempted to understand what users understand of trusting autonomous vehicles through cognitive processes and driving behaviors in a simulator study by enactment from users.

Design:
Subjects:36 drivers among which 24 men, 12 women-between 20 and 57 years of age and all of them having at least 1 year of driving experience. The mean subject was 36.4 years of age and had 7.08 years of experience.
- The driving scenarios were simulated using OpenDS software. Three test scenarios:
-Approaching a traffic light which is due to turn amber shortly.
-To overtake by another car.
-To meet a truck traveling at a slow pace.
-The participants drove the car in three conditions: manual, standard AV, and personalized AV, and the performances concerning speed and acceleration were compared.
  
  Apparatus:
The driving simulator incorporated an original car seat, steering wheel, pedals, and a 360° projection installation.Audio and video data were recorded for further analysis.
 
 Design:
The study was of within-subject design in which all subjects experienced all three driving conditions: manual, standard AV, and personalized AV. Driving performance and subjective ratings were collected.

Results:
Trust and user experience between the experimental driving style of the AV and manual and standardized AVs have been compared. Key findings include the following:

1. Only in the case of manual and customized modes was it found to have significant correlation with speed and acceleration, while the standardized AVs moved at constant speeds.
2. The comfort ratings were higher than the comfort baseline of 3.63 both for standard AV (4.6) and for personalized AV (5) with no difference between standard and personalized modes.
Manual driving had a slightly higher level of situational awareness than both standardized and personalized AVs, though there were not statistically significant differences among the three conditions-55.5 for respective conditions, 53.7, and 51.4.

Post-interview revealed that all the evaluation parameters that made people accelerate or move fast were noted by more than 70 percent of the respondents. Although most are open to sharing real-time driving data, there is privacy concern because only 40 percent agreed to share history about the use of the mobile phone.

\subsection{The Future of HCI in Autonomous Vehicles}

\subsubsection{ Innovation of Personalization }
This refers mainly to personalization within human-computer interaction in an autonomous vehicle, where the user experience is tailored to a particular individual's needs, preferences, and behavior\cite{41}. Improvements are driven in this direction by improvements in AI, machine learning, and data analytics, which help vehicles learn from users over time.Personalized interfaces allow seat height as well as seat settings, climate settings, and other infotainment settings to be set based on user profiles\cite{42}. Voice recognition units can perceive and act based on commands as well as the personal preferences of the driver, thereby providing overall improvement in the experience. There might be even personalization regarding the kind of route and type of driving suited to a user's preference for comfort, given the highest speed, or really nice views. With these technologies,the interfaces will begin to feel ever more intuitive and involving; hence, the autonomously run vehicle will indeed become an extension of personal space for the user\cite{43}.
\subsubsection{Future Prospects of AI and HCI in Autonomous Vehicles}
According to the future of artificial intelligence and human-computer interaction in autonomous vehicles, it promises a lot of change in personal transportation\cite{44}. Most importantly, it includes using interaction development to generate advanced user interfaces to be used for smooth communication between passengers and systems integrated into the vehicle. These rising levels of autonomous-ness within a vehicle dictate its intuitive interface, primarily needed in connection with user engagement, characterized by the utmost level of minimal distraction, by the use of touch-less controls, voice commands, and even gesture recognition\cite{45}.Of course, AI will make adaptive systems that learn about people's behavior and preferences. This would mean that one can have autonomy self-adjust to its comfort style of driving in order to predict whether it needs a smooth ride or more dynamic for traveling passengers based on historic data. There will also be contextual awareness due to the fact that AI would now know to read out the environment, such as traffic conditions, weather, and even the mood of the passengers, in order to make the vehicle adapt to its operations. For example, if a car had noticed that the passengers were very tired, it may offer them a ride in quiet or otherwise suggest where they should rest.
In addition, personalization will trickle down to every layer of the experience of the user. So, through its control systems or infotainment systems, the vehicle will be able to create a personal profile that determines settings as varied as climate control and seating positions, all the way down to media preferences\cite{46}. Those systems are going to optimize comfort, but in doing so, they will also allow choices in tailored entertainment, making suggestions about the content based on previous behaviors.
HCI will facilitate collaborative interactions of vehicles with other connected devices in positioning autonomous vehicles within smart transportation ecosystems\cite{47}. For instance, a car will talk to smart city infrastructure for optimal routing to reduce travel time while enhancing the user experience.Again, with the advancement in HCI design, it would be of utmost importance that the relationship between both parties be secure and trustworthy. The transparent feedback mechanism would bring the passenger on par with the decisions of the vehicle and its further maneuvers, along with the safety features involved, thus providing them with a sense of security and control even when the vehicle is working in the autonomous mode\cite{48}. This synergy of AI and HCI will go ahead and further redefine personal transport in the years to come, bringing safety, efficiency, and greater accessibility.

\section{\textbf{Conclusion}}
 Personalization of user interfaces for autonomous vehicles is an important frontier area in the domain of automotive innovation and human-computer interaction\cite{49}. It presents a number of specific challenges that require rather large-scale cognitive modeling of driver profiles, which has to ensure both trust and control by each user when interacting with an autonomous vehicle. This trust would be fundamental for this technology to become commonly accepted and can be achieved by customizing interfaces according to user preferences, driving behavior, and on-the-fly decision processes.

With personal user interfaces, intelligent transportation systems in future smart cities like Society 5.0 will increasingly rely on autonomous vehicles. These interfaces will make the driving experience enjoyable by taking complex tasks like lane changes and delegating them to automation, but they will also make aspects of the trip-from climate control to infotainment systems-focus on the personal preferences of users. All this will make such vehicles intuitive, friendly, and interesting to use, so it would be easy to adopt them in your daily life.Corporate labs like Honda have been at the forefront of personalizing autonomous vehicle interfaces, yet incorporating safety perspectives in complex scenarios as well as balancing such demands for safety remains a problem\cite{50}. Another concern also came to the fore-to-wit: data privacy and security.
In the future, it is highly promising because AI, IoT, and sensors allow vehicles to better identify and address user needs as they develop without compromise on unprecedented, unparalleled personalization.

Personalized interfaces with autonomous vehicles will define future transportation, improve the user experience, and ensure safety. However, despite the current challenges still facing research and development in such products, one can expect a lot of innovation to come forth. And as this field goes forward, we will witness products that not only capture people's imagination but also drive market value. Vehicles are going to be personal, smart components of daily life for any citizen living in Society 5.0. Human interaction with transportation is going to be transformed as vehicles become personal assistants while traveling.

\end{document}